\documentclass[preprint,prd,nofootinbib,tightenlines,amsmath]{revtex4}
\usepackage{epsfig}
\usepackage{graphicx}

\def\CM{{\cal M}}
\newcommand{\mP}{{\bar M}_{P}}

\begin{document}

\baselineskip=15pt


\hspace*{\fill} $\hphantom{-}$


\title{Discriminating between technicolor and warped extra dimensional model via pp $\to$ ZZ channel}

\author{O.~Antipin}
\email{oleg.a.antipin@jyu.fi}
\author{K.~Tuominen}
\email{kimmo.tuominen@phys.jyu.fi}
\affiliation{Department of Physics, University of Jyv\"askyl\"a, P.O.Box 35, FIN-40014 Jyv\"askyl\"a, Finland \\
and 
Helsinki Institute of Physics, P.O.Box 64, FIN-00014 University of Helsinki, Finland\\}

\date{January 28, 2009}

\begin{abstract} 
\vspace{2mm}
We explore the possibility to discriminate between certain strongly-coupled technicolor (TC) models and warped extra-dimensional models where the Standard Model fields are propagating in the extra dimension. We consider a generic QCD-like TC model with running coupling as well as two TC models with walking dynamics. We argue that due to the different production mechanisms for the lowest-lying composite tensor state in these TC theories compared to the first Kaluza-Klein graviton mode of warped extra-dimensional case, it is possible to distinguish between these models based on the angular analysis of the reconstructed longitudinal Z bosons in the $pp \to ZZ \to $ four charged leptons channel. 
\end{abstract}

\pacs{PACS numbers: }

\maketitle

\section {Introduction}

A clear signal of the New Physics (NP) that lies beyond the Standard Model (SM) would be the existence of heavy new particles or
``resonances''. In TeV mass range, the main arena of search for these objects will soon be the CERN LHC experiments as these particles can be produced or exchanged in high energy collisions. The direct experimental evidence for a resonance is the peak in the energy dependence of the measured cross sections seen above the SM background.

Once the resonance signal is observed, further analysis is needed to distinguish between scenarios that potentially may cause this effect. First step in this analysis would be the determination of the spin of the resonance which provides an important selection among different classes of
non-standard interactions. In the second step, then, one should test how likely this signal is accounted for by a certain NP model based on the 
parameter space of the model, particle spectrum and production mechanism(s) for the observed resonance.

In this regard, suppose LHC will observe a resonance in the $\sim$ 1.5-2 TeV mass range in the ZZ invariant mass distribution in the pp $\to$ ZZ $\to$ 4 charged leptons mode. Now, what could be the underlying theory that is responsible for this resonance? 

As a first possibility, recall \cite{Agashe:2007zd} that the first Kaluza-Klein (KK) mode of the spin-2 graviton from Randall-Sundrum (RS) extra-dimensional model \cite{Randall:1999ee} may well account for this enhancement. The reason is that for graviton with mass in the range $\sim$ 1.5-2 TeV, the graviton production via gluon fusion with subsequent decay to longitudinal Z boson pair is significantly dominating over all other possible sources within the RS model. In particular, the Vector Boson Fusion (VBF) production mode is found to be about an order of magnitude smaller for this mass range \cite{Agashe:2007zd}. However, it is hard to reach the graviton mass bigger than $\sim$ 2 TeV as the graviton cross-section is small and the irreducible SM background to pp $\to$ ZZ is starting to dominate for higher graviton masses.

Another spin-2 candidate for the enhancement in the $\sim$ 1.5-2 TeV mass range in this mode is the tensor bound state of some strongly-coupled 
Technicolor (TC) theory \cite{Weinberg:1979bn,Susskind:1978ms} produced via VBF and decaying to the longitudinal Z boson pair. Such spin-2 resonance plays also an important role in these TC models since it helps to delay unitarity violation in the longitudinal $W^{+} W^{-}\to ZZ$ and $ZZ\to ZZ$ scatterings. In these scatterings, spin-1 isospin-1 (technirho) vector resonance will appear only in the $t$ and $u$-channels in the $W^{+} W^{-}\to ZZ$ scattering and will not appear in the $ZZ \to ZZ$ scattering at all. Also, in the TC theories we will consider, the lightest spin-0 resonance is expected to be lighter than $\sim$ 1 TeV \cite{Hong:2004td}. Finally, due to the parity conservation, the axial resonance cannot directly participate in these tree-level scatterings either. This leaves the lightest spin-2 TC bound state as a strong candidate to account for the
assumed-to-be-observed resonance. 

Thus, in this paper we start with the assumption that there is indeed a spin-2 resonance observed and refer to other studies for discrimination techniques between spin-0, spin-1 and spin-2 resonances \cite{Allanach:2000nr,Allanach:2002gn,Cousins:2005pq,Osland:2008sy}. We will argue that due to the different production modes for the spin-2 states in RS and TC theories, it is possible to distinguish these models based on the angular analysis of the reconstructed longitudinal Z bosons. 

We will consider three different TC scenarios: Minimal Walking TC (MWT) \cite{Sannino:2004qp,Dietrich:2005jn,Dietrich:2006cm,Foadi:2007ue}, 
next to MWT (NMWT) \cite{Sannino:2004qp,Dietrich:2005jn,Dietrich:2006cm,Foadi:2007ue} and QCD-like TC theory. 
First two models have received significant attention in the literature due to the fact that these models successfully pass the 
electroweak precision tests (EWPT) \cite{Sannino:2008ha}. In addition, MWT leads to new candidates for cold dark matter \cite{Kouvaris:2007iq,Kainulainen:2006wq,Khlopov:2008ty,Kouvaris:2008hc} and, with addition of modest amount of TC-neutral matter, also to a novel unification of the SM couplings \cite{Gudnason:2006yj}. In all of these TC theories, the technimatter will be singlet under QCD and, thus, the VBF will be the dominant production mechanism for the spin-2 bound state. Parameters of the TC theories will be chosen such as to preserve the unitarity up to at least ( M$_{{\textrm{spin-2}}}^{{\textrm{TC}}}$ + 0.4 TeV ) and to account for the experimental evidence of the resonance. Once this is accomplished, we will mimic the same resonance in the RS model and compare the angular distributions of reconstructed Z bosons from both theories. Due to the different production mechanisms the angular distributions will be different and we will see that, in a long run, LHC will have a fair chance to conclusively distinguish between spin-2 states from TC and RS theories.

The paper is organized as follows. In section II we discuss how the different production mechanisms for the spin-2 state reveal themselves in the angular distributions of the longitudinal Z bosons. In section III we present the amplitudes for the $W^{+} W^{-}\to ZZ$ and $ZZ \to ZZ$ scatterings and specify the parameters values for the three TC models we consider based on the unitarity and consideration of modified Weinberg Sum Rules (WSR). In parallel with section III, section IV deals with the amplitudes and the parameters for the RS model. In section V we present our numerical results relevant for LHC phenomenology. Final discussions and conclusions are presented in section VI.

\section {Angular analysis}
\label{angular}

In this section we show how the angular distribution of the reconstructed Z bosons is connected to the production mechanism for the spin-2 resonance. 
Such resonance has the total of five possible polarization states, but gluons can produce only $|J J_Z\rangle=|2, \pm2\rangle$ and $|J J_Z\rangle=|2, 0\rangle$ since gluons do not have longitudinal polarizations and the total angular momentum has to be equal to $J=2$ (where we have chosen beam axis to be in the z-direction). Now, suppose that the two gauge bosons from the decay of the resonance are produced at the polar angle $\theta$. We rotate the gluon-produced resonance state specified by polarization tensor $\epsilon_{\mu\nu}(J, J_Z)$ by this angle \cite{Chung:1971ri,Antipin:2008hj}:

\begin{equation}
\epsilon_{\mu\nu}(2,J_Z)=\sum_{J_Z^{\prime}} D^{(J)*}_{J_Z J_Z^{\prime}}(0,\theta,0) \epsilon^{\prime}_{\mu\nu}(2,J_Z^{\prime}),
\label{rotation}
\end{equation}
where $\epsilon^\prime_{\mu\nu}(J,J_Z^{\prime})$ corresponds to the resonance state with the z-axis aligned with 
the direction of the decay products, and  $D^{(J)}_{J_Z J_Z^{\prime}}(\alpha,\theta,\gamma)\equiv \langle JJ_Z^{\prime}|R(\alpha,\theta,\gamma)|JJ_Z\rangle=
e^{-iJ_Z^{\prime}\alpha}d^{(J)}_{J_Z J_Z^{\prime}}(\theta)e^{-iJ_Z\gamma}$ is the familiar Wigner $D$-matrix. 
Independent Wigner small $d$-matrix elements for the spin-2 state are presented in the Appendix for completeness, see also \cite{Berman:1965gi}. Now we may easily derive the angular dependence of the helicity amplitudes for the channels appearing in our analysis. They follow from Eq.(\ref{rotation}) for the $|2, \pm2\rangle$ resonance state which is produced either by two $|1, 1\rangle$ or by two $|1, -1\rangle$ gluon states:

\begin{equation}
\epsilon_{\mu\nu}(2,\pm2)= d^{(2)}_{\pm 2 0}(\theta)\epsilon^{\prime}_{\mu\nu}(2,0)+ d^{(2)}_{\pm 2 1}(\theta)\epsilon^{\prime}_{\mu\nu}(2,1)+ d^{(2)}_{\pm 2 -1}(\theta)\epsilon^{\prime}_{\mu\nu}(2,-1).
\label{decomp}
\end{equation}

Now just use Clebsch-Gordan decomposition of the $\epsilon^{\prime}_{\mu\nu}(2,0)$ 
and $\epsilon^{\prime}_{\mu\nu}(2,\pm1)$ states in terms of 1$\otimes$1 final 
spin states to observe that, for example, helicity amplitude 
${\mathcal{A}}[g(\lambda_1)g(\lambda_2)\to Z(\lambda_3)Z(\lambda_4)]\equiv 
{\mathcal{A}}_{\lambda_1 \lambda_2 \lambda_3 \lambda_4}$ for  
${\mathcal{A}}_{+-00}\sim d^{(2)}_{2 0}(\cos\theta)$, ${\mathcal{A}}_{+-0-}\sim d^{(2)}_{2 1}(\cos\theta)$, 
and ${\mathcal{A}}_{+-0+}\sim d^{(2)}_{2 -1}(\cos\theta)$. Notice that we have not included 
the $d^{(2)}_{\pm 2 2}(\theta)\epsilon^{\prime}_{\mu\nu}(2,2)$ 
and $d^{(2)}_{\pm 2 -2}(\theta)\epsilon^{\prime}_{\mu\nu}(2,-2)$ terms in Eq.(\ref{decomp}) 
as the Z bosons from the decay of the spin-2 resonance have longitudinal polarization and, thus, these terms cannot contribute.
The $\epsilon_{\mu\nu}(2, 0)$ state of the resonance does not contribute due to the fact that a gluon is massless. 

For longitudinal VBF, the only possible mode for the produced resonance is $\epsilon_{\mu\nu}(2, 0)$ and the corresponding amplitude behaves as
${\mathcal{A}}_{0000}\sim d^{(2)}_{0 0}(\cos\theta)$.

Inherent to our analysis is the assumption that the spin-2 resonance is produced essentially at rest so that its decay products are mostly back to back. We assume that the center of mass frame of the resonance can be determined experimentally in the clean pp $\to$ ZZ $\to$ 4 charged leptons mode where the decay chain can be fully reconstructed.

\section{Technicolor models}

The original idea of technicolor \cite{Weinberg:1979bn,Susskind:1978ms} is the postulate of the existence of additional TC 
gauge interaction similar to QCD with the dynamical scale of the order of the electroweak scale and some new matter fields called technifermions taken to transform according to a suitable representation of the TC gauge group. One of the explicit TC models we will consider in this paper will be the "scaled-up" version of QCD with technifermions in the fundamental representation of the TC gauge group. In two other models, technifermions will be in the adjoint (MWT) and two-index symmetric (NMWT) representations of the TC gauge groups SU(2) and SU(3), respectively. Furthermore, in all these models the technifermions are taken to be singlet under QCD and assigned anomaly free $U_Y(1)$ hypercharges. The electroweak gauge group $SU(2)_L\times U_Y(1)$ is a part of the new global symmetry group associated with chiral dynamics of the technifermions. When strong dynamics of TC forces spontaneously breaks these new global symmetries the weak gauge bosons obtain their masses through the Higgs mechanism in usual manner while the photon of electromagnetism remains massless.

\subsection {Scattering of longitudinally polarized vector bosons}

We begin with the brief review of the formalism for the scattering of the (techni)pions. At high energies, the scattering of longitudinally polarized vector bosons ($V_L$) can be approximated by the scattering of the would-be-Goldstone bosons $W^a$. If we think of these Goldstone fields in analogy with the pions of QCD, we expect the $V_LV_L$ scattering amplitudes to be unitarized by a spin-one, isospin-one vector resonance, like the
techni-rho. As another alternative, if we think of the Goldstone fields in terms of the linear sigma model, we expect the scattering
amplitudes to be unitarized by a spin-zero, isospin-zero scalar field like the Higgs boson. Both of these states, together 
with the spin-2 resonance will be included when we consider unitarity of the scattering amplitudes.

We are interested in the strongly interacting longitudinal $V_L$ bosons in the TeV region. We will ignore the up-down fermion mass splittings and, therefore, the SU(2) ``isospin'' is conserved.  The $V_LV_L$ scattering amplitudes can then be written in terms of isospin amplitudes with the following assignment of the isospin indices,
\begin{equation}
W^a_L\ W^b_L\ \to\  W^c_L\ W^d_L\ ,
\label{threeseven}
\end{equation}
where $W_L$ denotes either $W_L^\pm$ or
$Z_L$, where $W_L^\pm=(1/\sqrt{2})(W_L^1 \mp iW_L^2)$ and
$Z_L=W_L^3$.  The scattering amplitude is given by
\begin{equation}
\CM(W^a_LW^b_L\to W^c_LW^d_L)\ =\ A(s,t,u)\delta^{ab}
\delta^{cd}\ +\ A(t,s,u)\delta^{ac}\delta^{bd}\ +
\ A(u,t,s)\delta^{ad}\delta^{bc}\ ,
\label{threeeight}
\end{equation}
where $a,b,c,d=1,2,3$, and $s$, $t$, and $u$ are the usual
Mandelstam variables. All the physics of $W_LW_L$ scattering is
contained in the amplitude function $A(s,t,u)$.

Given the amplitude function, the physical amplitudes for
boson-boson scattering are given as follows \cite{Golden:1995xv},
\begin{eqnarray}
\CM(W^+_LW^-_L\to Z_LZ_L)\ & = &\ A(s,t,u) \nonumber \\
\CM(Z_LZ_L\to W^+_LW^-_L)\ & = &\ A(s,t,u) \nonumber \\
\CM(W^+_LW^-_L\to W^+_LW^-_L)\ & = &\ A(s,t,u)\ +\ A(t,s,u) \nonumber \\
\CM(Z_LZ_L\to Z_LZ_L)\ & = &\ A(s,t,u)\
+\ A(t,s,u)\ +\ A(u,t,s) \nonumber \\
\CM(W^\pm_LZ_L\to W^\pm_LZ_L)\ & = &\ A(t,s,u) \nonumber \\
\CM(W^\pm_LW^\pm_L\to W^\pm_LW^\pm_L)\ & = &\ A(t,s,u)\ +
\ A(u,t,s)\ .
\label{threenine}
\end{eqnarray}

We will need only $\CM(W^+_LW^-_L\to Z_LZ_L)$ and $\CM(Z_LZ_L\to Z_LZ_L)$ amplitudes.

\subsection{Amplitude function and Unitarity}

Contribution of the spin-1 (technirho) and spin-0 (composite Higgs) resonances to the scattering is \cite{Foadi:2008xj}: 

\begin{equation}
A(s,t,u)=\left(\frac{1}{F_\pi^2}-\frac{3g_{V\pi\pi}^2}{M_V^2}\right)s
-\frac{h^2}{M_H^2}\frac{s^2}{s-M_H^2}
-g_{V\pi\pi}^2\left[\frac{s-u}{t-M_V^2}+\frac{s-t}{u-M_V^2}\right] \ ,
\label{eq:inv_2}
\end{equation}

Here $F_\pi=246$ GeV appropriate for the strong dynamics at the electroweak scale. Scaling up from QCD, note that the mass of the vector resonance should be as large as (246 GeV/93 MeV)$\times$770 MeV $\simeq$ 2 TeV. However in a theory with walking dynamics the resonances are expected to be lighter than in a running, QCD-like, setup. Similarly the coupling $g_{V\pi\pi}$ can be estimated by recalling that the QCD value that follows from $\Gamma(\rho\to \pi\pi)\simeq 150$~MeV would be $g_{V\pi\pi}\simeq 5.6$.

The contribution of a spin-two meson $F_2$ to the invariant amplitude comes from the effective chiral interaction Lagrangian between the traceless symmetric tensor nonet field $T_{\mu\nu}$ and the pions 
\cite{Foadi:2008xj,Sannino:1995ik,Harada:1995dc}, 
$L_T=-g_2 F_{\pi}^2 Tr[T_{\mu\nu}p^{\mu}p^{\nu}]=
-\frac{g_2}{\sqrt{2}} (F_2)_{\mu\nu}\partial^{\mu} \vec{\pi}\cdot \partial^{\nu} \vec{\pi}+\cdots$, and the resulting contribution to the amplitude is 
\begin{eqnarray}
\label{tensor}
A_2(s,t,u)=\frac{g_2^2}{2(M_{F_2}^2-s)}\left[-\frac{s^2}{3}+\frac{t^2+u^2}{2}\right]-\frac{g_2^2 s^3}{12 M_{F_2}^4} \ ,
\end{eqnarray}
where $M_{F_2}$ and $g_2$ are the mass of the spin-2 meson and its coupling with the pions, respectively. Again, one can obtain a reference value 
for $g_2$ by relating to QCD where $m_{f_2}\simeq 1275$~MeV and $\Gamma(f_2\to \pi\pi) \simeq 160$~MeV gives $|g_2| 
\simeq 13$~GeV$^{-1}$ so that $|g_2|F_\pi \simeq 1.2$. Scaling the dimensionful numbers up to the eletroweak scale results in $|g_2| \simeq 4$~TeV$^{-1}$. It is easy to check that in the $m_{W,Z}=0$ limit the $\left[-\frac{s^2}{3}+\frac{t^2+u^2}{2}\right]$ piece 
$\sim d^{(2)}_{0 0}(\cos\theta)$ in accordance with the angular analysis  section \ref{angular}. We note that the tensor interaction $L_T$ above can, on the gravity side, be compared to the gauge-invariant Lagrangian for the KK graviton $H_{\mu\nu}$ field coupled to the scalar bosons which is 
$L_G\sim(H^{\mu\nu}-\frac{1}{2}g^{\mu\nu}H) D_{\mu}^\dagger \Phi D_{\nu}\Phi$, where $H\equiv H^{\mu}_{\mu}$ \cite{Han:1998sg}. The appearance of the  $\sim s^3$ term in Eq. (\ref{tensor}), which would not be present in the graviton case, is due to the absence of the $\sim g^{\mu\nu}H$ term, {\em i.e.} the tracelessness condition for the $T_{\mu\nu}$.

To constrain the parameter space by unitarity of the $\pi\pi$ scattering we apply the analysis of \cite{Foadi:2008xj} which we briefly review here for completeness. For the unitarity analysis the most general amplitude should be expanded in its isospin $I$ and spin $J$ components, $a^I_J$, but it turns out that the $I=0$ $J=0$ component,
\begin{equation}
a_0^0(s) = \frac{1}{64\pi} \int_{-1}^1 d\cos\theta \left[3A(s,t,u)+A(t,s,u)+A(u,t,s)\right] \ ,
\end{equation}
has the worst high energy behavior, and is therefore taken as a basis for the analysis. 

The possible resonances which contribute to the unitarity of $\pi\pi$ scattering here are a light Higgs, an axial (which will be lighter than the vector in two of the three TC theories), vector and the tensor. The axial can only be lighter than the vector in a Walking Technicolor theory (WT), 
where the second Weinberg Sum Rule is modified. Moreover, the chances of the axial being lighter than the vector are increased as the conformal window is approached, and the $S$ parameter decreases. Also, light composite Higgs can naturally emerge in strongly coupled theories with matter in higher dimensional representations such as MWT and NMWT \cite{Hong:2004td}.

The axial resonance cannot directly participate in the tree-level exchanges in the $\pi\pi$ scattering due to parity invariance. Rather, the constraint due to $A^a_\mu$ field is indirect and appears in the $\pi\pi$ scattering because the pion eaten by the $W$ boson contains a certain amount of the longitudinal component of $A^a_\mu$. As a consequence, the $g_{V\pi\pi}$ and $h$ coupling are affected by the presence of $A^a_\mu$, but since the dependence on $M_A$ comes together with other new parameters, it turns out that $g_{V\pi\pi}$ and $h$ remain completely free to take on any value. To see the effects of a light axial resonance in the $\pi\pi$ scattering one imposes the WSR's and simultaneously requires that the $S$ parameter is small. Since the WSR's are affected by walking dynamics the resulting constraints on the allowed region in the $(M_V,g_{V\pi\pi})$ space are different with walking dynamics than in a theory with a QCD-like dynamics and a heavy axial.

The WSR's, appropriately modified to account for the walking dynamics \cite{Foadi:2007ue}, read
\begin{eqnarray}
& & S=4\pi\left[\frac{F_V^2}{M_V^2}-\frac{F_A^2}{M_A^2}\right] \ , \label{eq:WSR0} \\
& & F_V^2-F_A^2=F_\pi^2 \label{eq:WSR1} \ , \\
& & F_V^2 M_V^2 - F_A^2 M_A^2 = a\frac{8\pi^2}{d(R)}F_\pi^4 \label{eq:WSR2} \ ,
\end{eqnarray}
where $F_V$ ($F_A$) and $M_V$ ($M_A$) are decay constant and mass of the vector (axial) resonance, $d(R)$ is the dimension of the fermion representation of the underlying gauge theory, and $a$ is an unknown number. In WT $a$ is expected to be positive and of order one while in a QCD-like theory $a=0$. For the MWT with two flavors in the adjoint representation of SU(2) we take $a/d(R)=1/3$, and the naive contribution to the $S$ parameter is $1/2\pi\simeq 0.15$. In the NMWT we have two flavors in the two-index symmetric representation of SU(3), $a/d(R)=1/6$, and the naive estimate of $S$ is $1/\pi\simeq 0.3$. We will also consider the constraints for a running theory, {\em i.e.} $a=0$.

Furthermore, in addition to the WSR's of Eqs.~(\ref{eq:WSR0}) - (\ref{eq:WSR2}) one demands that vector and axial resonances should not be broad by excluding the regions in the parameter space corresponding to,
\begin{eqnarray}
\Gamma_V/M_V < 1/2 \ , \quad \Gamma_A/M_A < 1/2 \ .
\label{eq:narrow}
\end{eqnarray}

Finally, combining the above constraints with the unitarity constraints gives the allowed regions 
for the parameters $\hspace{1mm}h$,$\hspace{1mm}M_V,\hspace{1mm}g_{V\pi\pi}$,$\hspace{1mm}M_{F_2}$ and
$\hspace{1mm}g_2$ in the three TC theories we will consider (see Appendix of \cite{Foadi:2008xj} where these constraints
were translated to the constraints on the values of $\hspace{1mm}M_V,\hspace{1mm}g_{V\pi\pi}$. In our analysis
we added the tensor mass and coupling as extra parameters while in \cite{Foadi:2008xj} only 
$\hspace{1mm}h$,$\hspace{1mm}M_V,\hspace{1mm}g_{V\pi\pi}$ were considered). Specific values for the parameters will be given in  detail in the section \ref{numerics}, where we present our numerical results. 

To conclude this section and to prepare for the numerical analysis of the total cross sections, we discuss the relevant production mechanism for spin-2 TC bound state, namely, VBF via $WW$ or $ZZ$ \cite{Agashe:2007zd}. The probability for emission of (an almost) collinear longitudinal $W/Z$ by a quark (or anti-quark) is suppressed by electroweak factor of $\sim \alpha_{ EW } / \left( 4 \pi \right)$. However, the coupling of longitudinal $W/Z$ to spin-2 TC tensor is enhanced. Moreover, VBF can proceed via valence quarks through the $u u$, $d d$ or $u d$ scatterings in addition to $u \bar{u}$ and $d \bar{d}$ annihilation (which are suppressed due to the smaller sea quark content).

Hence, the other amplitudes with transverse polarizations for initial or final state bosons can be neglected due to the smaller couplings to the spin-2 TC tensor. As was explained in sec. \ref{angular}, the expected angular dependence of the matrix element is $ {\cal{M}}_{ 0 0 0 0 } \sim d^{(2)}_{0 0}(\cos\theta)\sim  \left(  1 - 3\cos^2 \theta \right)$ as we will also see to arise from our numerical calculation.

The required parton-level cross-section is given by
\begin{eqnarray}
\frac{ d \hat{ \sigma } \left( V_L V_L
\rightarrow ZZ \right) }{ d \cos \hat{ \theta } }
& \approx & \frac{ | {\cal M}_{ 0 0 0 0 } |^2 }{ 64 \pi \hat{s} }
\end{eqnarray}
where the subscript $L$ on $V$ denotes longitudinal polarization.

The probability distribution for a quark of energy $E$ to emit a longitudinally polarized gauge boson of energy $xE$ and transverse momentum $p_T$ (relative to quark momentum) is approximated by \cite{Han:2005mu}:
\begin{eqnarray}
\frac{ d P^L_{ V / f } \left( x, \; p_T^2\right) }{ d p_T^2 }
& = & \frac{ g_V^2 + g_A^2 }{ 4 \pi^2 } \frac{ 1 - x }{x}
\frac{ ( 1 - x ) M_V^2 }{ \Big[ p_T^2 + ( 1 - x )
M_V^2 \Big]^2 }
\nonumber \\
\label{Lemission}
\end{eqnarray}

For an example of $WW$ fusion from $u d$ scattering, the proton-level cross-section can then be written as
\begin{eqnarray}
\sigma \left( pp \rightarrow ZZ
\right) & \ni &
 \int d x_1 d x_2 d x_1^W d x_2^W d p_{ T \; 1 }^2
d p_{ T \; 2 }^2 \nonumber \\
& \times &
\frac{ d P^L_{ W/u }
\left( x_1^W, \; p_{ T \; 1 }^2 \right) }{ d p_{ T \; 1 } ^2 }
\frac{ d P^L_{ W/d }
\left( x_2^W, \; p^2_{ T \; 2 } \right) }{ d p_{ T \; 2 } ^2 }
\nonumber \\
& \times& f_u (x_1, Q^2 ) f_d (x_2, Q^2 ) \hat{ \sigma } \left( \hat{s} \right)
 \nonumber \\
 &+& (u \leftrightarrow d)\nonumber\\
& \approx &
\int d x_1 d x_2 d x_1^W d x_2^W
f_u (x_1, Q^2 ) f_d (x_2, Q^2 )
\nonumber \\
& \times& P^L_{ W/u }
\left( x_1^W \right) P^L_{ W/d }
\left( x_1^W \right)
\hat{ \sigma } \left( s x_1 x_2 x^W_1 x^W_2 \right)\nonumber \\
&+& (u \leftrightarrow d)
\nonumber \\
\end{eqnarray}
where in the second line, we have used the fact that from Eq. (\ref{Lemission}) the
average $p^2_T$ of the longitudinal $V$ is given by $\sim ( 1 - x ) M^2_V \ll \left( x^W_{ 1, \; 2} E \right)^2$.
Here, $x^W_{ 1, \; 2 } E \sim m^G_n \sim$ TeV is roughly the energy of the longitudinal $V$
in order to produce an {\em on}-shell composite tensor state. Hence, we can neglect transverse momenta in the parton-level cross-section, i.e., set
$\hat{s} \approx s x_1 x_2 x^W_1 x^W_2$ and integrate over the transverse momenta to obtain
total probabilities, $P^L_{ W/d } (x) = P^L_{ W/ u } (x) \approx g^2 / \left( 16 \pi^2 \right) \times (1 - x) / x$.
Also, $f_{ u, d }$ is the parton distribution function (PDF) corresponding to $u$, $d$ quarks in the proton;
the $u$ quark (or $W^+$) can come from the first proton and
$d$ quark (or $W^-$) from the second proton or vice versa.
Expressions for contributions from $W_L/Z_L$ emission from various
other combinations of quarks and anti-quarks inside the protons
can be similarly obtained. Finally, we mention that the VBF offers the possibility to tag 
two additional highly energetic forward jets \cite{Han:2005mu}. In this
case we have to consider the $ZZ$ + 2 jets SM background. Additional complication is the 
fact that in the present configuration of LHC experiments, pseudorapidity range $\sim$ 6-10 is not covered.
We will not pursue the forward-jets-tagging option here and only require two (almost) back-to-back Z bosons in the central region
with "central-jet-vetoing" \cite{Barger:1990py}. \\

\section{RS model}
\label{RS}

In this section, we briefly review the RS model and present the amplitudes needed for the numerical analysis.
The RS framework \cite{Randall:1999ee} features an extra-dimension which is taken to be a slice of AdS$_5$ space.
At the endpoints of this five-dimensional space ($\phi=0,\pi$) reside two branes which are usually labeled as an ultraviolet (UV) Planck brane and an IR (TeV) brane. The large hierarchy of scales is resolved by a geometrical exponential factor. Due to the warped geometry, the relationship between the $5D$ mass scales (taken to be of order $\mP$ where $\mP$ is the reduced Planck mass in four dimensions) and those in an effective $4D$ description depend on the field-localization in the extra dimension. The zero-mode of the graviton is localized near the UV brane which has a Planckian fundamental scale, whereas the Higgs sector is localized near the IR brane where  the "red-shifted" fundamental scale is in the TeV-range. Postulating modest-sized $5^{{\textrm{th}}}$ dimension with radius $R$ and curvature $k$ the ratio of the TeV and Planck scales is $\sim e^{-k\pi R}$ and can be numerically obtained by setting $kR$ $\approx$11. 

The SM fields are allowed to propagate in the extra dimension \cite{Davoudiasl:1999tf,Pomarol:1999ad,Grossman:1999ra,Huber:2000ie,Gherghetta:2000qt}.
In this scenario there are KK excitations of SM gauge and fermion fields in addition to those of the graviton. These states have masses in the TeV range and are localized near the TeV brane. The SM particles are the zero-modes of the 5D fields, and the profile of a SM fermion in the extra dimension depends on its 5D mass. By localizing light fermions near the Planck brane and heavier ones near the TeV brane, the contributions to the FCNC and EWPT are suppressed by factors TeV/$\Lambda\ll 1$ where $\Lambda\sim{\mathcal{O}}(\mP)$. As a consequence, the KK graviton whose profile is peaked at the TeV brane will couple mostly to the top quark, Higgs (or, by equivalence theorem, to the longitudinal $W$ and $Z$ bosons), and KK excitations of the SM fields \cite{Agashe:2007zd,Antipin:2008hj,Fitzpatrick:2007qr,Davoudiasl:1999jd}. Standard Model gluons have a flat profile so that their coupling to KK graviton is suppressed only by a factor of the size of the extra dimension (in units of radius of curvature), i.e., $k \pi R\approx 35$. 

Now, we consider the couplings relevant for the production and decay. The production is dominated by gluon fusion and the coupling of gluons
to KK gravitons is given by \cite{Agashe:2007zd,Davoudiasl:2000wi}:
\begin{eqnarray}
C^{ A A G }_{ 0 0 n } & = &
\frac{e^{ k \pi R }}{\mP}
\frac{ 2 \left[ 1 - J_0 \left( x_n^G \right) \right] }{ k \pi R
\left( x_n^G \right)^2 | J_2 \left( x_n^G \right) | }
\end{eqnarray}
where $J_{ 0, 2 }$ denote Bessel functions and the values $x^G_n = 3.83, 7.02, 10.17, 13.32$ give masses of the first four KK gravitons: $m^G_n = k e^{ - k \pi R } x^G_n$. Also, in scenarios where the Higgs field is localized at the TeV brane, the coupling of the graviton to the longitudinal Z bosons is equal to $e^{ k \pi R }/  \mP$.

The relevant matrix elements for the process $gg \rightarrow ZZ$, via KK graviton are \cite{Agashe:2007zd}:
\begin{equation}
{\cal M}^G_{ \lambda_1 \lambda_2 \lambda_3 \lambda_4 }
\left( g^a g^b \rightarrow ZZ \right) =
- C^{ A A G }_{ 0 0 n }
\left( \frac{ x^G_n c }{ m^G_n } \right)
%
\times \sum_n \frac{ \delta_{ a b }\,
[{\cal A}_{ \lambda_1 \lambda_2 \lambda_3
\lambda_4 }]}{ \hat{s} - m_n^2 + i
\Gamma_G m_n }
\label{MG1}
\end{equation}
where $\lambda_i$ refer to initial and final state polarizations, c$\equiv k/ \mP$ and $a, b$ are color factors,
\begin{eqnarray}
\Gamma_G & = & \frac{ 20 (c\, x^G_n)^2 \, m^G_n }{ 960 \pi}
\label{width}
\end{eqnarray}
is the total decay width of KK graviton in our treatment and we have used
$\mP\, e^{ - k \pi R } = m^G_n/(x^G_n c)$. For c=1, the width of the graviton is $\approx 10 \%\ $ 
of the graviton mass  which will be the input for our numerical analysis later. This width value is higher than 
the one obtained in Ref.\cite{Agashe:2007zd} but see \cite{Antipin:2008hj}, where additional graviton decay modes
to the $W_{KK}(Z_{KK})W_L(Z_L)$ were considered. We have
\begin{eqnarray}
{\cal A}_{ + + 0 0 } & = & {\cal A}_{ - - 0 0 } = 0
 \\
{\cal A}_{ + - 0 0 } & = & {\cal A}_{ - + 0 0 } =
\frac{
\left( 1 - 1 / \beta_Z^2 \right)
\left( \beta_Z^2 - 2 \right)
\Big[ \left( \hat{t} - \hat{u} \right)^2
- \beta_Z^2 \hat{s}^2 \Big] \hat{s} }{ 8 M_Z^2 } \nonumber
\nonumber
\label{ggfusion}
\end{eqnarray}
where $\beta_Z^2 = 1 - 4 M_V^2 / \hat{s}$ and the hatted variables are in the parton center of mass frame. 
The other amplitudes with transverse polarizations for $Z$ bosons (i.e., $\lambda_{3,4} = +, -$) can be neglected since these are suppressed relative
to the above by $\sim \log \left( \mP / \hbox{TeV} \right)$. As we showed in section \ref{angular} and as now can be seen also explicitly, 
${\cal A}_{ + - 0 0 } \to -\sin^2\hat{\theta} \, \hat{s}^2/2 \sim d^{(2)}_{2 0}(\cos\theta)$ as $\beta_Z \to 1$.

Similarly as in the case of technicolor, to reliably compare against the possible experimental signal we need the parton-level signal ($V=Z$) cross-section, averaged over initial state spins and colors. This is given by \cite{Agashe:2007zd}:
\begin{eqnarray}
\frac{ d \hat{ \sigma } \left( g g \rightarrow ZZ \right) }{ d \cos
\hat{ \theta } }
& \approx &
\frac{| {\cal M}_{ +- 00 } |^2 }{ 1024 \pi \hat{s} }
\label{dsig/dc}
\end{eqnarray}
where a factor of $1/2$ has been included for identical bosons in the final state, initial helicity averaging has been accounted for by a factor of $1/4$
and a factor of $1/8$ accounts for color averaging. Note that ${\cal M}_{ +- 00 }$ is the only independent non-zero matrix element for the above process.
The total parton level cross section $\hat{\sigma}$ is related to the proton-level total signal cross-section as usual:
\begin{eqnarray}
\sigma ( pp \rightarrow ZZ )
\!\!& = &
\!\!\!\int \!\!dx_1 dx_2 f_g \left( x_1, Q^2 \right)\!f_g \left( x_2, Q^2 \right)
\!\hat{ \sigma } \left(
x_1 x_2 s \right), \nonumber \\
\label{sigtot}
\end{eqnarray}
where $f_g$ are the gluon PDF's and $Q^2 \sim ({m^G_n})^2$ is the typical momentum transfer in the partonic process for resonant production
of a KK graviton.

\section{Numerical results}

In this section we consider our predictions for the LHC. We evaluate numerically the total proton-level cross-sections as defined in previous sections. To begin with, we consider the SM background and then move on to present our numerical results.

\subsection{SM background}

We concentrate on the leptonic decay mode of the two $Z$ bosons ($ZZ\rightarrow 4 \ell$),
based on considerations of the backgrounds.
The irreducible background to the $ZZ$ final state, i.e., SM contribution
to $pp \rightarrow ZZ + X$ is dominated by $q \bar{q}$ annihilation:
gluon fusion is very small in the SM since it proceeds via a fermion loop. Hence, the interference
of KK graviton signal (dominated by $gg$) with the SM background is negligible. We also checked that $ZZ$ + 2 jets SM background
and its interference with the TC signal, with the requirements specified in the
previous sections, are negligible. The parton-level cross-section, averaged
over quark colors and spins is given by \cite{Eichten:1986eq}
\begin{equation}
\frac{ d \hat{ \sigma } \left( q_i \bar{q}_i \rightarrow ZZ \right) }
{ d \hat{t} }  =  \frac{ \pi \alpha^2 \left( L_i^4 + R_i^4 \right) }
{ 96 \sin^4 \theta_W \cos^4 \theta_W \hat{s}^2 } \times
\Big[ \frac{ \hat{t} }{ \hat{u} } + \frac{ \hat{u} }{ \hat{t} } +
\frac{ 4 M_Z^2 \hat{s} }{ \hat{t} \hat{u}}
- M_Z^4 \left( \frac{1}{ \hat{t}^2 } + \frac{1}{ \hat{u}^2 } \right) \Big],
\nonumber \\
\end{equation}
where $L_u = 1 - 4/3 \sin^2 \theta_W$, $R_u = - 4/3 \sin^2 \theta_W$,
$L_d= -1 + 2/3 \sin^2 \theta_W$ and $R_d = + 2/3 \sin^2 \theta_W$
This cross-section exhibits forward/backward peaking due to $t/u$ channel exchange, whereas
the resonant KK graviton and TC signals do not have this feature. Hence, a cut on pseudo-rapidity $\eta$ is useful to reduce this background keeping the
signal (almost) unchanged. The total proton-level cross section is obtained as in the RS model case, namely, using Eq. (\ref{sigtot}).

Finally, we will exploit the fact that the SM background is dominated by transversely
polarized Z bosons. Thus, if we boost to the Z rest frame and make an appropriate cut on
the angle between the charged lepton momenta and the direction of the boost we select more
data with longitudinally polarized Z bosons \cite{Park:2001vk}.

\subsection{Total cross sections: TC vs RS}
\label{numerics}

We can now present the comparison between TC and RS models. For numerical simulations, we checked the SM background with CALCHEP program \cite{Pukhov:2004ca} and used Mathematica program for graviton and TC signals.  CTEQ5LO PDF's were exploited throughout (in their Mathematica distribution package \cite{Pumplin:2002vw} as well as intrinsically called by CALCHEP). 

\begin{figure}[htb]
\centering
\includegraphics[width=3.0in]{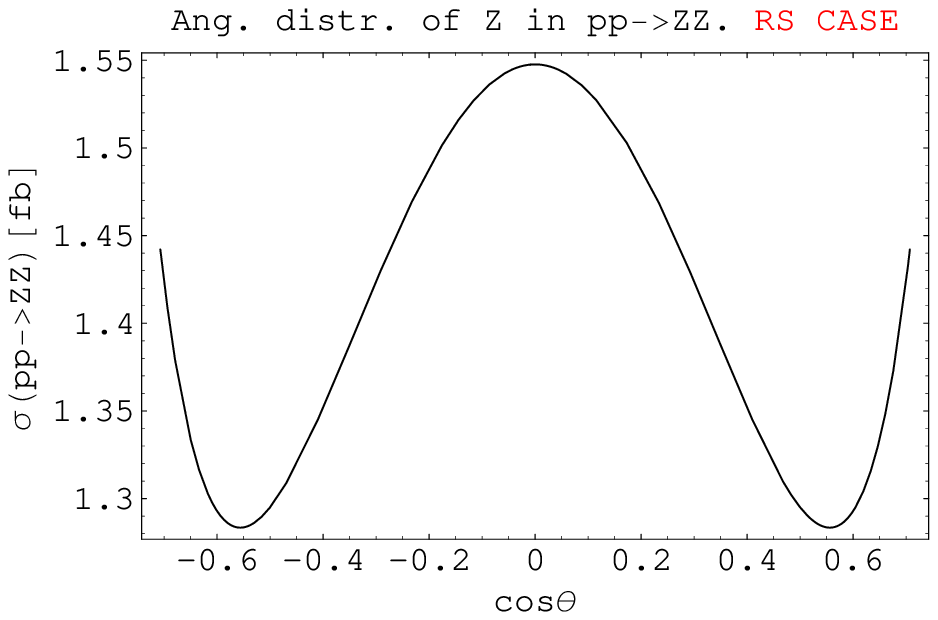} \includegraphics[width=3.in]{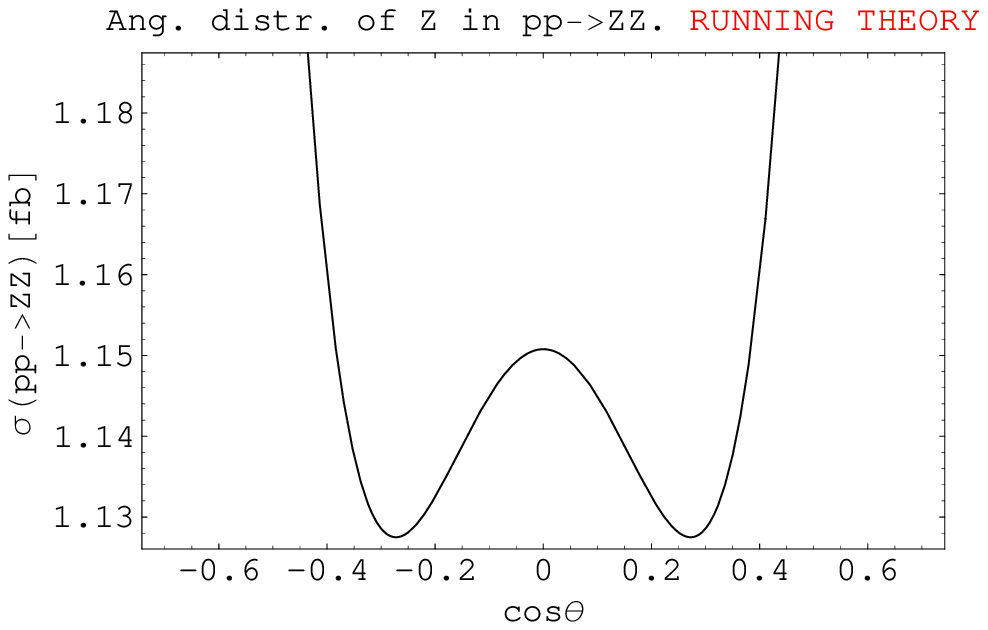}\\
\hspace{7mm}(a)\hspace{72mm}(b)
\caption{The total (signal + background) angular distributions of the reconstructed 
Z bosons in \\pp $\to$ ZZ mode integrated in three-resonance-width region 
(a) RS case (b) QCD-like TC theory case. }
\label{RS2}
\end{figure}

Since the main difference in angular distribution between RS and TC spin-2 signals is in the central rapidity region and also since SM background has forward/backward peaking we impose hard pseudorapidity cuts which we will specify below. Also, throughout the analysis we fix $\hspace{1mm}\Gamma_G$=0.1m$_1^{G}$,
$\hspace{1mm}\Gamma_{TC}$=0.1$M_{F_2}$, c$\hspace{1mm}\equiv k/ \mP$=1  in rough agreement with the expectations from the RS and scaled-up QCD estimates. We neglect the W and Z boson masses which is a reasonable approximation to the accuracy of our final result. Given the above values, $M_{F_2}$ remains the only parameter in the RS theory. Our
strategy is then, for any given mass of the spin two resonance, to find a set of allowed parameters for Technicolor best mimicking the RS-case. Depending on the value of $M_{F_2}$ this corresponds to either QCD-like or walking Technicolor as follows:

{\bf Running theory}: $M_{F_2}$=2 TeV, $\hspace{2mm} g_2=9.5 \times 10^{-3}$ GeV$^{-1}$,
$\hspace{2mm}$h=0.6, $\hspace{2mm}m_h$=130 GeV, $\hspace{2mm}M_V=1.6$ TeV, 
$\hspace{2mm}g_{V\pi\pi}$=2.26. Now, we integrate over three-resonance-width region. The results are shown in Fig.\ref{RS2} and in 
Table.\ref{tableZZ}. We have $\sim$ 20 total events in the $\eta<0.88$ region. Also, the angular dependence due to the presence of the spin-2 resonances shown in Fig.\ref{RS2} is given by $\sin^4\theta$ and $(1 - 3 \cos^2\theta)^2$ in RS and TC cases, respectively.
We included the $Z \to \tau \tau$ channel into statistics of Table.\ref{tableZZ} due to the fact that the $\sim$500 GeV energy $\tau$'s from $Z$ decay will have a decay length of $\ell=\gamma \tau c\approx 20$ mm and therefore might leave visible tracks in the detector \cite{Bengtsson:1985ym,Antipin:2007pi}.

\begin{figure}[htb]
\centering
\includegraphics[width=3.0in]{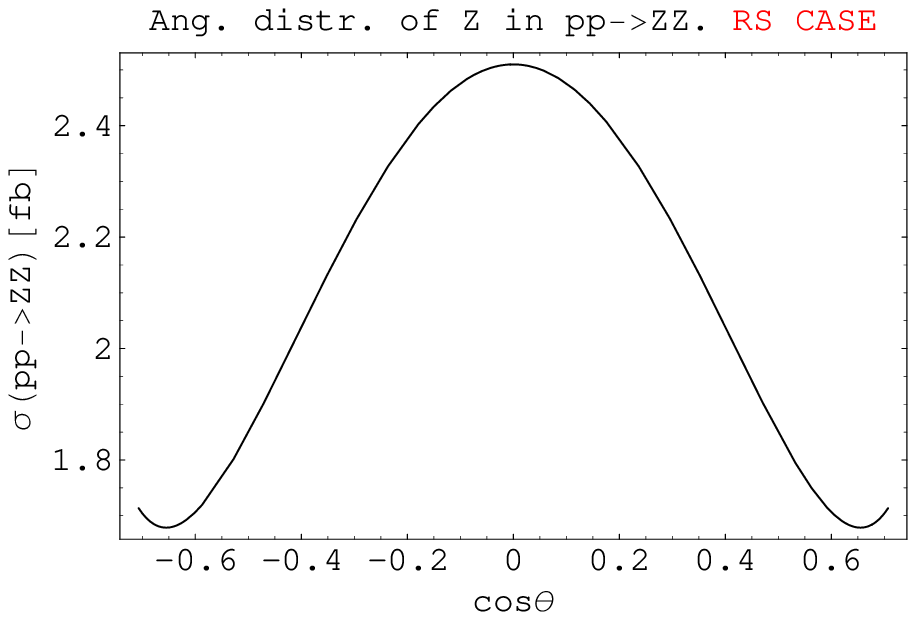} \includegraphics[width=3.in]{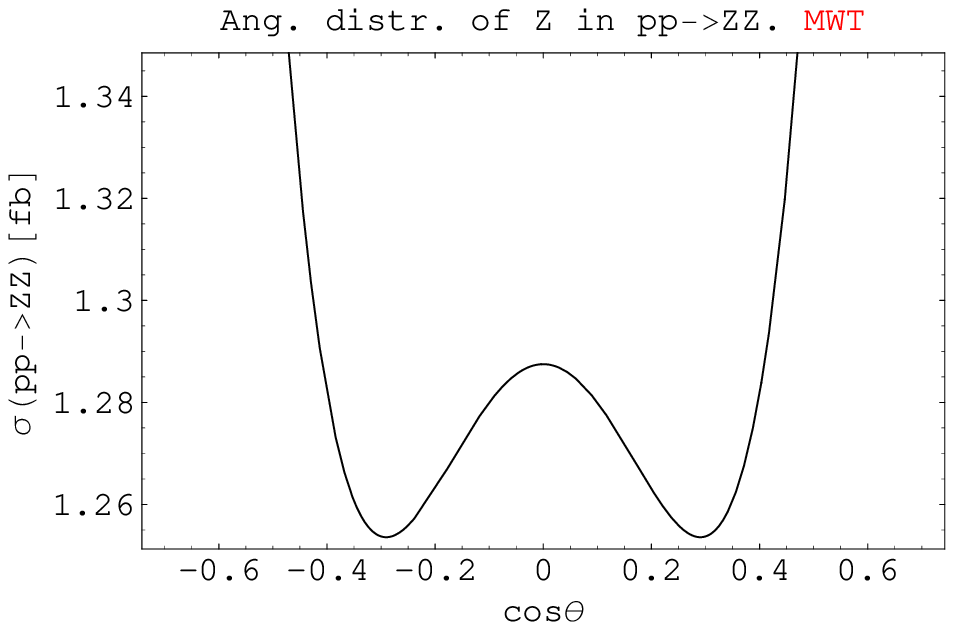}\\
\hspace{7mm}(a)\hspace{74mm}(b)
\caption{The total (signal + background) angular distributions of the reconstructed 
Z bosons in \\pp $\to$ ZZ mode integrated in three-resonance-width region 
(a) RS case (b) MWT theory case.}
\label{RS3}
\end{figure}

{\bf MWT}: $M_{F_2}$=1.8 TeV,$\hspace{2mm} g_2=9.5 \times 10^{-3}$ GeV$^{-1}$, $\hspace{2mm}$h=0.57, $\hspace{2mm}m_h$=130 GeV, $\hspace{2mm}M_V=1.2$ TeV, $\hspace{2mm}g_{V\pi\pi}$=2.0. This parameter space corresponds to the region where axial is lighter than the vector. We integrated over three-resonance-widhts region and made the cut $|\cos\theta|<$0.7 on the angle $\theta$ between the charged lepton momenta and the boost direction of the Z boson. The results are shown in Fig.\ref{RS3} and in Table.\ref{tableZZ}. Now, we have 29 and 19 total events for RS and TC  cases respectively in the $\eta<0.88$ region. The spin-2 and spin-1 mass values for this case were also intended to mimic the corresponding RS prediction $m_1^{G}\approx$1.5 $m_1^{KK}$, where $m_1^{KK}$ is the first spin-1 KK mass \cite{Davoudiasl:2000wi}.  

\begin{figure}[htb]
\centering
\includegraphics[width=3.0in]{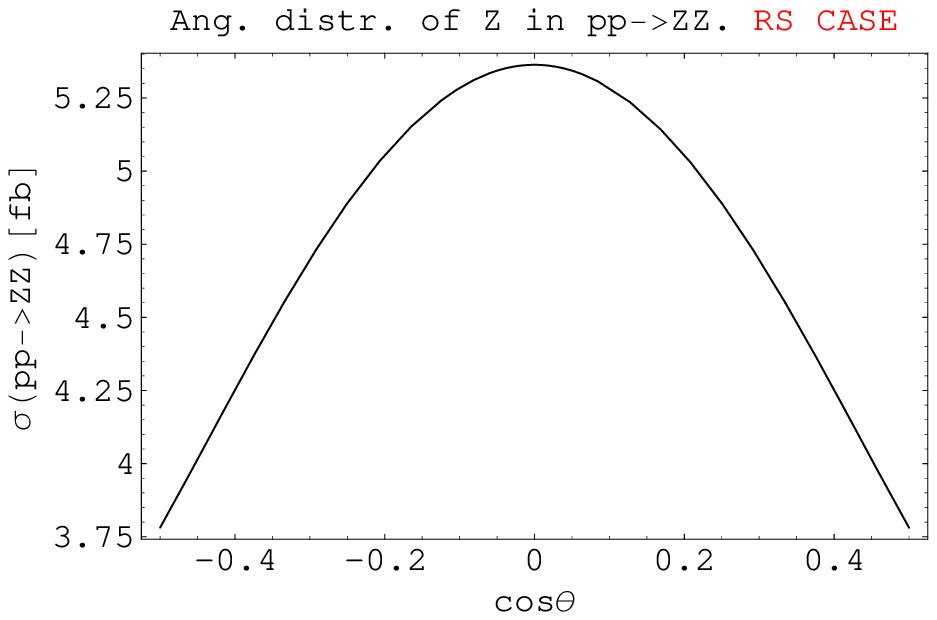} \includegraphics[width=3.in]{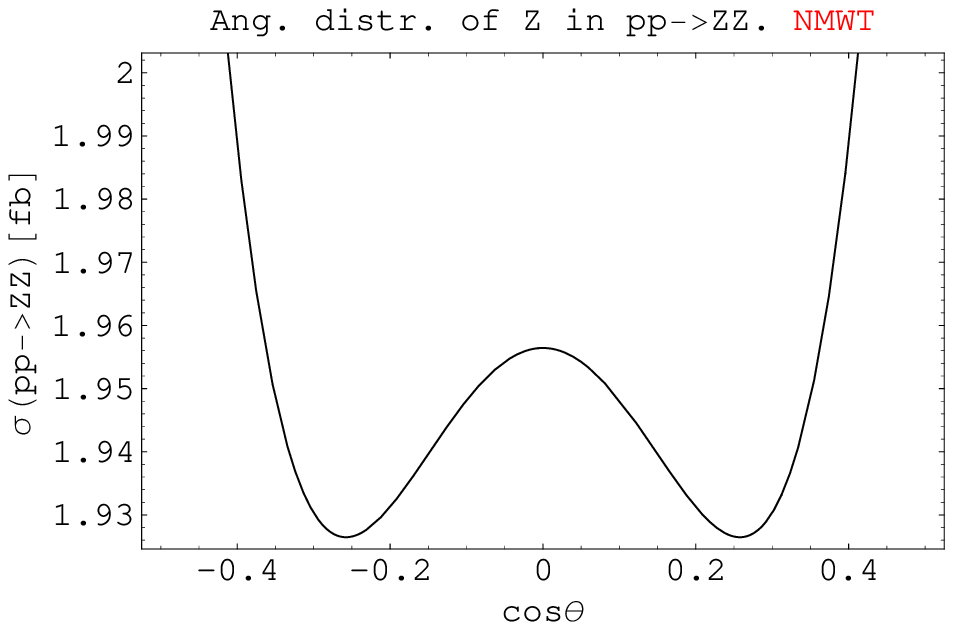}\\
\hspace{7mm}(a)\hspace{74mm}(b)
\caption{The total (signal + background) angular distributions of the reconstructed 
Z bosons in\\pp $\to$ ZZ mode integrated in three-resonance-width region 
(a) RS case (b) NMWT theory case.}
\label{RS4}
\end{figure}

{\bf NMWT}: $M_{F_2}$=1.6 TeV,$\hspace{2mm} g_2=9.5 \times 10^{-3}$ GeV$^{-1}$,
$\hspace{2mm}$h=0.55, $\hspace{2mm}m_h$=130 GeV, $\hspace{2mm}M_V=0.89$ TeV, 
$\hspace{2mm}g_{V\pi\pi}$=1.8. This parameter space also corresponds to the region where axial is lighter
than the vector. We integrated over three-resonance-width region and also took $\eta<0.55$.
In addition, the cut $|\cos\theta|<$0.7  on the angle $\theta$ between the charged lepton momenta and the boost direction of the Z boson was again applied. The results are shown in Table.\ref{tableZZ} and in Fig.\ref{RS4}. In this case, 
we have $\sim$ 48 and 19 total events in the $\eta<0.55$ region for RS and TC cases, respectively.

These numerical values demonstrate the generic problem that lowering the spin-2 tensor mass, RS production 
dominates more and more over corresponding TC and SM background cases because gluon PDF's are larger in this
region as compared to quark ones. Thus, it is very hard to mimic the production for both sides with low spin-2 mass. Thus, if the LHC happens to 
observe spin-2 state with such a low mass, RS would be the favored explanation out of the possibilities considered here. 
One might think that it is possible to lower the RS production by decreasing the value of c$\equiv k/ \mP$ parameter. But then from Eq. (\ref{width}) the graviton width decreases significantly. Then the width of the RS graviton and spin-2 TC bound states are not expected to be the same anymore and therefore this should help to distinguish them in experiment. We cannot lower the spin-2 TC resonance width because for such a big value of the tensor coupling in our numerical analysis the width of the spin-2 TC bound state is expected to get bigger, not smaller, than in the guiding scaled-up-QCD case.
The fact that in our analysis we kept width of the spin-2 TC resonance close to the expected one from QCD but significantly increased the tensor coupling to the pions provides another tension in this problem. 

We emphasize that the above dominance of RS vs TC spin two production is characteristic to the RS model described in Section \ref{RS}. In particular, the ultraviolet scale in this model was taken to be $\sim \mP$ but other possibilities can be envisioned. Among these possibilities, in \cite{Davoudiasl:2008hx} volume-truncated version, by a factor y $\approx$ 6, of the RS model was proposed with an ultraviolet cut off scale $O (10^3)$ TeV. In this "Little RS (LRS)" model, the graviton cross-section considered in our paper would be lowered by a factor of $O(y)$ and thus could agree in magnitue with the corresponding TC case with low spin-2 mass better \cite{Davoudiasl:2008hx}. However, the differences in the angular distributions between RS and TC remain robust.


\begin{table}
\caption{Signal $pp \to ZZ \to$ 4 charged leptons ($e$,$\mu$,$\tau$) cross-section $\sigma$(in fb) 
with the corresponding leading SM background. Numbers correspond to $\eta<0.88$ cut case 
(except the NMWT case where $\eta<0.55$).
We assume 100 $\%\ $ efficiency for our clean 4-lepton signal.  }
\label{tableZZ}
\begin{center}
\begin{tabular}{|c|c|c|c|c|}
\hline
Running Theory:$m_{spin-2}^{RS,TC}$=2 TeV&$\sigma$(fb)&$\#$ of events/1000 fb$^{-1}$&S/B&S/$\sqrt{B}$\\
\hline
Signal $G_1$ $\to$ ZZ $\to$ 4 lept.  &0.012&12&1.5&4.2\\
\hline
Signal $G_{TC}$ $\to$ ZZ $\to$ 4 lept.  &0.010&10&1.3&3.5\\
\hline
SM ZZ$\to$ 4 lept. &0.008&8&&\\
\hline
\hline

\hline
MWT:$m_{spin-2}^{RS,TC}$=1.8 TeV&$\sigma$(fb)&$\#$ of events/1000 fb$^{-1}$&S/B&S/$\sqrt{B}$\\
\hline
Signal $G_1$ $\to$ ZZ $\to$ 4 lept.  &0.021&21&2.6&7.4\\
\hline
Signal $G_{TC}$ $\to$ ZZ $\to$ 4 lept.  &0.011&11&1.4&3.9\\
\hline
SM ZZ$\to$ 4 lept. &0.008&8&&\\
\hline
\hline
NMWT:$m_{spin-2}^{RS,TC}$=1.6 TeV&$\sigma$(fb)&$\#$ of events/1000 fb$^{-1}$&S/B&S/$\sqrt{B}$\\
\hline
Signal $G_1$ $\to$ ZZ $\to$ 4 lept.  &0.040&40&5.0&14.1\\
\hline
Signal $G_{TC}$ $\to$ ZZ $\to$ 4 lept.  &0.011&11&1.4&3.9\\
\hline
SM ZZ$\to$ 4 lept. &0.008&8&&\\
\hline
\end{tabular}
\end{center}
\end{table}

We want to stress that once we assume that the 1.5-2 TeV resonance is observed in experiment, it means that the tensor
has to dominate in TC case under consideration which in turn implies that due to different production mechanism there will be a difference between RS and TC cases. Of course, it is possible that in TC case parameters are different than the ones we considered but then it is not always possible to find resonant behavior because spin-2 is \emph{the only} resonance in our treatment which can appear in the s-channel (except the composite Higgs which is expected to be light) and so the tensor coupling should be large enough (QCD reference value would be appropriate choice, for example). The problem for such a QCD-like value of the tensor coupling is the dominance of the SM background, which is about twice bigger than the TC signal. Summarizing, we choose the TC parameters based on the a) requirement of unitarity, b) saturation of Weinberg Sum Rules and c) experimental observation of a spin-2 resonance.

Also, we would like to point out that in the typical RS model we are considering, the ratio between the graviton coupling
to gluons and the graviton coupling to longitudinal gauge bosons is $\sim 1/k\pi R \approx$ 1/35 and, thus, one cannot 
change the relative size of the gluon fusion vs. VBF graviton productions mechanisms. Therefore, the VBF mechanism 
remains subleading to the gluon fusion even when, similar to TC case, one tries to increase the graviton coupling to "would-be" Goldstones (this would, however, be possible by increasing the c$\hspace{1mm}\equiv k/ \mP$ value). 

\section {Conclusions}

We analyzed three TC models and showed that it is possible to find parameter space for them which leads to spin-2 
production similar to the RS case for the same mass and width of the tensor resonance. We emphasize that the choices 
of the parameter values we made for the TC theory are rather uniquely fixed by the requirement of unitarity and the WSR.
The only unnatural value was the tensor coupling constant $g_2$ which was somewhat large compared to the QCD-like theory
predictions. The most natural region of the spin-2 mass where the two models can yield similar results was found to be about 2 TeV. For smaller mass, RS production is expected to dominate due to high gluon densities and for higher mass the SM background becomes dominant. We showed that still 
(assuming the two models give similar signals) it is possible to discriminate the models based on angular analysis after we know the mass, width, and S/B values from experiment. Knowing these values will suggest the corresponding angular distributions for the two theories. 

We note that our analysis can be also used as a prediction for the TC theories only (without the reference to the RS). In this case, perhaps
the tensor coupling value should be taken smaller (close to the QCD value $g_2 \sim $4 TeV$^{-1}$). Still, in the resonant region,
the tensor will dominate as long as the other parameters are taking values similar to the ones we used in our analysis. 
Angular distributions for this case will be dictated by the S/B ratio.

\begin{acknowledgments}
We thank Francesco Sannino and Amarjit Soni for a careful reading of the manuscript and many fruitful discussions. 
\end{acknowledgments}

\appendix
\section{Spin-2 Wigner small d-matrix}
\begin{eqnarray}
d^{(2)}_{22}(\beta)&=&\frac{(1+\cos\beta)^2}{4},\hspace{5mm} d^{(2)}_{21}(\beta)=-\frac{1+\cos\beta}{2}\sin\beta, \hspace{5mm} d^{(2)}_{2-1}(\beta)=-\frac{1-\cos\beta}{2}\sin\beta\nonumber\\
d^{(2)}_{20}(\beta)&=&\frac{\sqrt{6}}{4}\sin^2\beta,\hspace{10mm} d^{(2)}_{2-2}(\beta)=\frac{(1-\cos\beta)^2}{4},\hspace{9mm} d^{(2)}_{10}(\beta)=-\sqrt{\frac{3}{2}}\sin\beta \hspace{1mm}\cos\beta \nonumber\\
d^{(2)}_{11}(\beta)&=&\frac{1+\cos\beta}{2}(2\cos\beta-1),\hspace{10mm} d^{(2)}_{1-1}(\beta)=\frac{1-\cos\beta}{2}(2\cos\beta+1),\nonumber\\
d^{(2)}_{00}(\beta)&=&\frac{3\cos^2\beta-1}{2}.
\label{Wigner}
\end{eqnarray}


\begin{thebibliography}{999}

\bibitem{Agashe:2007zd}
  K.~Agashe, H.~Davoudiasl, G.~Perez and A.~Soni,
  Phys.\ Rev.\  D {\bf 76}, 036006 (2007)
  [arXiv:hep-ph/0701186].

\bibitem{Randall:1999ee}
  L.~Randall and R.~Sundrum,
  Phys.\ Rev.\ Lett.\  {\bf 83}, 3370 (1999)
  [arXiv:hep-ph/9905221].

\bibitem{Weinberg:1979bn}
  S.~Weinberg,
  Phys.\ Rev.\  D {\bf 19}, 1277 (1979).

\bibitem{Susskind:1978ms}
  L.~Susskind,
  Phys.\ Rev.\  D {\bf 20}, 2619 (1979).

\bibitem{Hong:2004td}
  D.~K.~Hong, S.~D.~H.~Hsu and F.~Sannino,
  Phys.\ Lett.\  B {\bf 597}, 89 (2004)
  [arXiv:hep-ph/0406200].

\bibitem{Allanach:2000nr}
  B.~C.~Allanach, K.~Odagiri, M.~A.~Parker and B.~R.~Webber,
  JHEP {\bf 0009}, 019 (2000)
  [arXiv:hep-ph/0006114].

\bibitem{Allanach:2002gn}
  B.~C.~Allanach, K.~Odagiri, M.~J.~Palmer, M.~A.~Parker, A.~Sabetfakhri and B.~R.~Webber,
  JHEP {\bf 0212}, 039 (2002)
  [arXiv:hep-ph/0211205].

\bibitem{Cousins:2005pq}
  R.~Cousins, J.~Mumford, J.~Tucker and V.~Valuev,
  JHEP {\bf 0511}, 046 (2005).

\bibitem{Osland:2008sy}
  P.~Osland, A.~A.~Pankov, N.~Paver and A.~V.~Tsytrinov,
  Phys.\ Rev.\  D {\bf 78}, 035008 (2008)
  [arXiv:0805.2734 [hep-ph]].

\bibitem{Sannino:2004qp}
  F.~Sannino and K.~Tuominen,
  Phys.\ Rev.\  D {\bf 71}, 051901 (2005)

\bibitem{Dietrich:2005jn}
  D.~D.~Dietrich, F.~Sannino and K.~Tuominen,
  Phys.\ Rev.\  D {\bf 72}, 055001 (2005)

\bibitem{Dietrich:2006cm}
  D.~D.~Dietrich and F.~Sannino,
  Phys.\ Rev.\  D {\bf 75}, 085018 (2007)

\bibitem{Foadi:2007ue}
  R.~Foadi, M.~T.~Frandsen, T.~A.~Ryttov and F.~Sannino,
  Phys.\ Rev.\  D {\bf 76}, 055005 (2007)

\bibitem{Sannino:2008ha}
  F.~Sannino,
  arXiv:0804.0182 [hep-ph].

\bibitem{Gudnason:2006mk}
  S.~B.~Gudnason, T.~A.~Ryttov and F.~Sannino,
  Phys.\ Rev.\  D {\bf 76}, 015005 (2007)
  [arXiv:hep-ph/0612230].

\bibitem{Gudnason:2006yj}
  S.~B.~Gudnason, C.~Kouvaris and F.~Sannino,
  Phys.\ Rev.\  D {\bf 74}, 095008 (2006)
  [arXiv:hep-ph/0608055].

\bibitem{Kouvaris:2007iq}
  C.~Kouvaris,
  Phys.\ Rev.\  D {\bf 76}, 015011 (2007)
  [arXiv:hep-ph/0703266].

\bibitem{Kainulainen:2006wq}
  K.~Kainulainen, K.~Tuominen and J.~Virkaj\"arvi,
  Phys.\ Rev.\  D {\bf 75}, 085003 (2007)
  [arXiv:hep-ph/0612247].

\bibitem{Khlopov:2008ty}
  M.~Y.~Khlopov and C.~Kouvaris,
  Phys.\ Rev.\  D {\bf 78}, 065040 (2008)
  [arXiv:0806.1191 [astro-ph]].

\bibitem{Kouvaris:2008hc}
  C.~Kouvaris,
  Phys.\ Rev.\  D {\bf 78}, 075024 (2008)
  [arXiv:0807.3124 [hep-ph]].

\bibitem{Chung:1971ri}
  S.~U.~Chung,
  ``SPIN FORMALISMS,''
  CERN-71-08; 
Lectures given in the Academic Training Program of CERN 1969-1970.


\bibitem{Antipin:2008hj}
  O.~Antipin and A.~Soni,
  JHEP {\bf 0810}, 018 (2008)
  [arXiv:0806.3427 [hep-ph]].

\bibitem{Berman:1965gi}
  S.~M.~Berman and M.~Jacob,
  Phys.\ Rev.\  {\bf 139}, B1023 (1965).

\bibitem{Golden:1995xv}
  M.~Golden, T.~Han and G.~Valencia,
  arXiv:hep-ph/9511206.

\bibitem{Foadi:2008xj}
  R.~Foadi, M.~Jarvinen and F.~Sannino,
  arXiv:0811.3719 [hep-ph].

\bibitem{Sannino:1995ik}
  F.~Sannino and J.~Schechter,
  Phys.\ Rev.\  D {\bf 52}, 96 (1995)
  [arXiv:hep-ph/9501417].

\bibitem{Harada:1995dc}
  M.~Harada, F.~Sannino and J.~Schechter,
  Phys.\ Rev.\  D {\bf 54}, 1991 (1996)
  [arXiv:hep-ph/9511335].

\bibitem{Han:1998sg}
  T.~Han, J.~D.~Lykken and R.~J.~Zhang,
  Phys.\ Rev.\  D {\bf 59}, 105006 (1999)
  [arXiv:hep-ph/9811350].



\bibitem{Davoudiasl:1999tf}
  H.~Davoudiasl, J.~L.~Hewett and T.~G.~Rizzo,
  Phys.\ Lett.\  B {\bf 473}, 43 (2000)
  [arXiv:hep-ph/9911262].

\bibitem{Pomarol:1999ad}
  A.~Pomarol,
  Phys.\ Lett.\  B {\bf 486}, 153 (2000)
  [arXiv:hep-ph/9911294].

\bibitem{Grossman:1999ra}
  Y.~Grossman and M.~Neubert,
  Phys.\ Lett.\  B {\bf 474}, 361 (2000)
  [arXiv:hep-ph/9912408].

\bibitem{Huber:2000ie}
  S.~J.~Huber and Q.~Shafi,
  Phys.\ Lett.\  B {\bf 498}, 256 (2001)
  [arXiv:hep-ph/0010195].

\bibitem{Gherghetta:2000qt}
  T.~Gherghetta and A.~Pomarol,
  Nucl.\ Phys.\  B {\bf 586}, 141 (2000)
  [arXiv:hep-ph/0003129].

\bibitem{Fitzpatrick:2007qr}
  A.~L.~Fitzpatrick, J.~Kaplan, L.~Randall and L.~T.~Wang,
  arXiv:hep-ph/0701150.

\bibitem{Davoudiasl:1999jd}
H.~Davoudiasl, J.~L.~Hewett and T.~G.~Rizzo,
Phys.\ Rev.\ Lett.\  {\bf 84}, 2080 (2000)
[arXiv:hep-ph/9909255].

\bibitem{Davoudiasl:2000wi}
  H.~Davoudiasl, J.~L.~Hewett and T.~G.~Rizzo,
  Phys.\ Rev.\  D {\bf 63}, 075004 (2001)
  [arXiv:hep-ph/0006041].

\bibitem{Han:2005mu}
  T.~Han,
  arXiv:hep-ph/0508097.

\bibitem{Barger:1990py}
  V.~D.~Barger, K.~m.~Cheung, T.~Han and R.~J.~N.~Phillips,
  Phys.\ Rev.\  D {\bf 42}, 3052 (1990).


\bibitem{Eichten:1986eq}
  E.~Eichten, I.~Hinchliffe, K.~D.~Lane and C.~Quigg,
  Phys.\ Rev.\  D {\bf 34}, 1547 (1986).


\bibitem{Park:2001vk}
  S.~C.~Park, H.~S.~Song and J.~H.~Song,
  Phys.\ Rev.\  D {\bf 65}, 075008 (2002)
  [arXiv:hep-ph/0103308].



\bibitem{Pukhov:2004ca}
  A.~Pukhov,
  arXiv:hep-ph/0412191.

\bibitem{Pumplin:2002vw}
  J.~Pumplin, D.~R.~Stump, J.~Huston, H.~L.~Lai, P.~Nadolsky and W.~K.~Tung,
  JHEP {\bf 0207}, 012 (2002)
  [arXiv:hep-ph/0201195].


\bibitem{Bengtsson:1985ym}
H.~U.~Bengtsson, W.~S.~Hou, A.~Soni and D.~H.~Stork,
  Phys.\ Rev.\ Lett.\  {\bf 55}, 2762 (1985).

\bibitem{Antipin:2007pi}
  O.~Antipin, D.~Atwood and A.~Soni,
  Phys.\ Lett.\  B {\bf 666}, 155 (2008)
  [arXiv:0711.3175 [hep-ph]].

\bibitem{Davoudiasl:2008hx}
  H.~Davoudiasl, G.~Perez and A.~Soni,
  Phys.\ Lett.\  B {\bf 665}, 67 (2008)
  [arXiv:0802.0203 [hep-ph]].

\end{thebibliography}
\end{document}